\documentclass{INTERSPEECH2023}

\interspeechcameraready 

\usepackage{pgfplots}
\pgfplotsset{compat = newest}
\usepackage{xcolor}
\usepackage{multirow}
\usepackage{tabularx}
\usepackage{colortbl}
\usepackage{floatrow}
\DeclareFloatFont{tiny}{\tiny}

\usepackage{pgfplotstable}
\newcommand{\app}{\raise.17ex\hbox{$\scriptstyle\sim$}}

\title{FlexiAST: Flexibility is What AST Needs}
\name{Jiu Feng*, Mehmet Hamza Erol*, Joon Son Chung, Arda Senocak \thanks{*These authors contributed equally to this work. This work was supported by the National Research Foundation of Korea (NRF) grant funded by the Korea government (MSIT) (No. RS-2023-00212845).}}
\address{
  Korea Advanced Institute of Science and Technology, South Korea}

\begin{document}

\maketitle

\begin{abstract}
The objective of this work is to give patch-size flexibility to Audio Spectrogram Transformers (AST). Recent advancements in ASTs have shown superior performance in various audio-based tasks. However, the performance of standard ASTs degrades drastically when evaluated using different patch sizes from that used during training. As a result, AST models are typically re-trained to accommodate changes in patch sizes. To overcome this limitation, this paper proposes a training procedure to provide flexibility to standard AST models without architectural changes, allowing them to work with various patch sizes at the inference stage - FlexiAST. This proposed training approach simply utilizes random patch size selection and resizing of patch and positional embedding weights. Our experiments show that FlexiAST gives similar performance to standard AST models while maintaining its evaluation ability at various patch sizes on different datasets for audio classification tasks.
\end{abstract}
\section{Introduction}
The latest developments in transformer models~\cite{vaswani2017attention} that rely purely on attention have had a significant impact in both audio processing~\cite{gong21b_interspeech,gong2022ssast,chen2022hts,koutini2021efficient,baade2022mae,huangmasked,chong2022masked,niizumi2022masked,nagrani2021attention,hsu2021hubert,Baevski2020wav2vec2A,zhu2023multiscale} and computer vision~\cite{dosovitskiyimage,Girdhar2019video,lu2019vilbert,Carion2020EndtoEndOD,caron2021dino,touvron2021training,zheng2021rethinking,wang2021pyramid,fan2021multiscale,li2022mvitv2} fields. When it comes to audio classification, all of the recent transformer-based approaches use a patch-based system. The input spectrograms are divided into fixed-size patches to create the tokens used for the transformer's inputs. While patches are an essential aspect of transformer models, prior approaches have primarily concentrated on the training paradigms for audio spectrogram transformers. The AST~\cite{gong21b_interspeech} adapts vision transformers to audio spectrograms and applies supervised learning with initialization from ImageNet~\cite{deng2009imagenet} pre-trained ViT~\cite{dosovitskiyimage, touvron2021training} models, while SSAST~\cite{gong2022ssast} extends it with self-supervision. The complexity of the transformer models increases drastically with the input sequence length. Thus, efficient ways of reducing the complexity is crucial. MAE-AST~\cite{baade2022mae} improves upon SSAST by utilizing a masked auto-encoder (MAE) approach to reduce the number of tokens. Additionally, several concurrent studies also explore some methods for optimizing the computational complexity of transformers through masking, patchout techniques and pooling, including~\cite{koutini2021efficient,huangmasked,chong2022masked,niizumi2022masked,zhu2023multiscale}.

\begin{figure}[ht!]
\centering
{
\resizebox{\linewidth}{!}{%
\begin{tabular}{c}
\includegraphics[width = 1.0\linewidth]{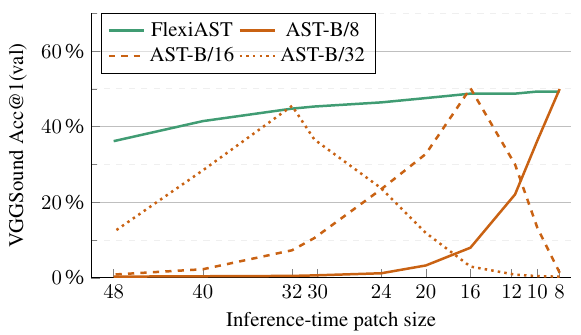} \\ 
\end{tabular}
}
}
\caption{\textbf{Standard ASTs vs. FlexiAST.} The performance of standard ASTs when evaluated on different patch sizes reveals their lack of flexibility.}
\label{fig:teaser}
\vspace{-4mm}
\end{figure}
Our main focus differs from other AST-based methods in that we prioritize the patch size of the audio spectrogram transformers (AST). While~\cite{gong21b_interspeech,baade2022mae} employ a fixed 16x16 patch size, audio-based tasks vary in their requirements and may necessitate different patch sizes to achieve optimal performance. This typically entails re-training the AST (with certain patch size) to accommodate changes in patch sizes, since standard ASTs (including Vision Transformers - ViT~\cite{dosovitskiyimage}) function best only at their trained patch size. Adjusting an already trained AST with bilinear interpolation to evaluate on different patch sizes results in a significant performance degradation as Figure~\ref{fig:teaser} displays. Thus, it is evident that standard ASTs lack flexibility.

This motivates us to design \emph{one single AST model} that can seamlessly work at different patch sizes without significant performance degradation while achieving comparable performance to standard ASTs trained at fixed patch sizes. With this objective in mind, we explore a learning process that can give flexibility to a standard AST model. The closest work to ours is FlexiViT~\cite{beyer2022flexivit}, which provides flexibility to Vision Transformers (ViT) to perform well at various patch sizes. We draw inspiration from the findings of this work and study the methodologies suitable for the flexibilization of standard ASTs. To accomplish this task, we take the following steps: (1) Instead of using a fixed patch size during training, we randomly select patch sizes (2) We resize the patch and positional embedding weights for the chosen patch size. The rest of the model (AST Transformer Encoder and Linear Layer) and training procedure remain identical to the standard AST, as depicted in Figure~\ref{fig:pipeline}. Additionally, we explore how various tasks necessitate distinct resizing methodologies. For instance, audio classification tasks concentrate on audio context, and, as such, resizing patch embeddings in the frequency axis together with the time axis may not alter the underlying semantic information. Conversely, in speaker identification tasks, frequency is directly linked to speaker-specific information. Hence, we have explored that FlexiAST requires patch embedding resizing solely along the time axis for speaker identification.

Our main contributions can be summarized as follows: (1) We demonstrate that standard ASTs are not flexible enough to be evaluated at patch sizes different from the size they were trained on; (2) We provide a training procedure that offers flexibility to standard ASTs without requiring architectural changes; (3) We show that this simple approach improves the flexibility of ASTs on different datasets, namely AudioSet~\cite{gemmeke2017audio}, VGGSound~\cite{VGGSound}, ESC-50~\cite{piczak2015esc}, Speech Commands~\cite{warden2018speech}, and VoxCeleb~\cite{nagrani2020voxceleb}.
\vspace{-4mm}\section{Approach}
\subsection{Preliminaries}
Audio Spectrogram Transformer (AST)~\cite{gong21b_interspeech} is a transformer based architecture that takes an audio spectrogram $x \in \mathbb{R}^{f \times t}$ as its input. It tokenizes $x$ into a sequence $x_i \in \mathbb{R}^{p \times p}$ of patches and computes the patch embeddings $e_i = (e_{i}^{k})_{k=1}^d \in \mathbb{R}^d$ for all the patches through $e_{i}^{k}=vec(x_i)^{T}vec(\omega_k)$, where $\omega=(\omega_k)_{k=1}^{d} \in \mathbb{R}^{d \times p \times p}$ are the patch embedding weights, and $vec$ function flattens a multi-dimensional array to a one-dimensional vector. Afterward, the trainable positional embeddings $\pi=(\pi_i)_{i=1}^{h \times w} \in \mathbb{R}^{(h \times w) \times d}$ are added to each of the patch embeddings $s_i=e_i+\pi_i$ for the final version of the tokens, where $h$ and $w$ represent the dimensions for the grid of tokens obtained. Finally, a {\fontfamily{pcr}\selectfont [CLS]} token is inserted at the beginning of the obtained token sequence, and the final sequence is fed into a multi-layered transformer encoder. The transformed output of the {\fontfamily{pcr}\selectfont [CLS]} token is used as the representation of the audio spectrogram for the downstream tasks. Since audio datasets typically do not contain large amounts of data required by transformers, AST initializes itself from a pretrained ViT weights to improve the training process and the performance.

\subsection{ASTs are not Flexible}
The patch embedding weights $\omega$ and positional embeddings $\pi$ are the only components of an AST with parameterizations depending on the patch size. Thus, in case a pretrained AST model is evaluated at a different patch size than the original parameterization (trained fixed patch size), these two components must be adjusted appropriately. More specifically, if we want to evaluate a pretrained standard AST with patch embedding weights $\omega \in \mathbb{R}^{d \times p \times p}$ and positional embeddings $\pi \in \mathbb{R}^{(h \times w) \times d}$ at the patch size $2p \times 2p$, then the patch embedding weights and positional embeddings should be resized into $\hat{\omega} \in \mathbb{R}^{d \times 2p \times 2p}$ and $\hat{\pi} \in \mathbb{R}^{(\lceil h/2 \rceil \times \lceil w/2 \rceil) \times d}$ respectively (assuming no overlap of patches). To achieve this in a most conventional and intuitive way, one can simply use bilinear interpolation operation as the original ViT paper~\cite{dosovitskiyimage} use this resizing method on the positional embeddings. However, Figure \ref{fig:teaser} illustrates that the performance of the AST models collapses when evaluated at different patch sizes with this resizing methodology. This implies that simply resizing the patch embedding weights of an AST in an intuitive way does not yield flexibility across multiple patch sizes. In the following sections, we propose a methodology to train flexible AST models that perform well across multiple patch sizes without a significant performance drop.
\begin{figure}[t!]
\centering
{
\resizebox{\linewidth}{!}{%
\begin{tabular}{c}
\includegraphics[width = 1.0\linewidth]{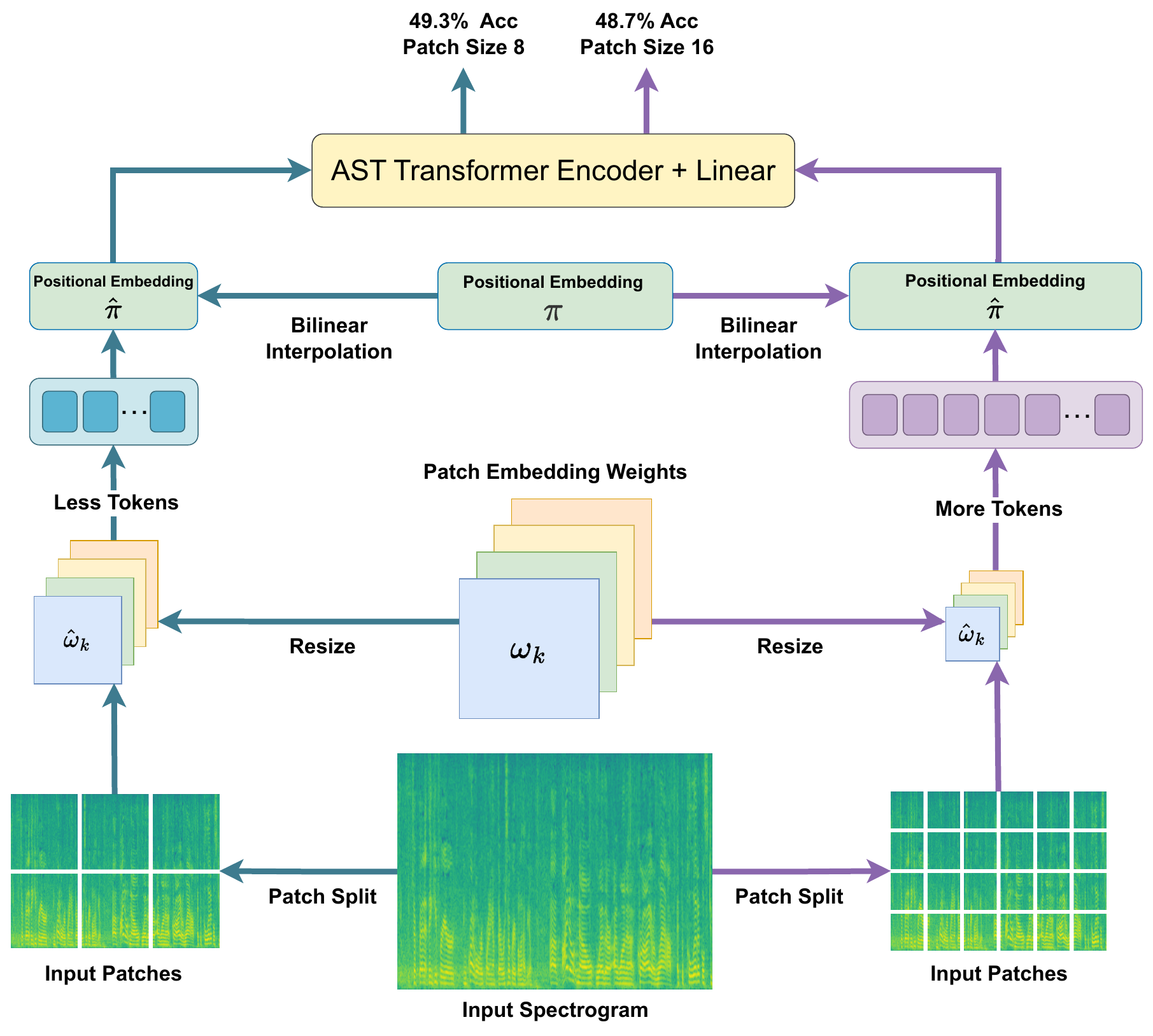} \\
\end{tabular}
}
}
\caption{\textbf{Our FlexiAST framework.} It is a standard AST model (architecturally identical) with random patch size selection and resizing operation on patch embeddings weights during training.}
\label{fig:pipeline}
\vspace{-2mm}
\end{figure}

\subsection{Approach and Training}
To achieve flexification, we only refine the training pipeline of standard ASTs in the light of FlexiViT's~\cite{beyer2022flexivit} findings, without making any architectural changes on standard ASTs. The new training pipeline involves two key stages: (1) selecting a random patch size during training as opposed to a fixed patch size, and (2) adjusting the weights of the patch embeddings based on the chosen patch size. We randomly select a patch size $\hat{p}$ during training from a set of patch sizes $P = \{8, 10, 12, 16, 20, 24, 30, 32, 40, 48\}$. Then, we resize the patch embedding weights and positional embeddings appropriately to match the selected patch size. 
The remaining training process is the same as standard ASTs, shown in Figure \ref{fig:pipeline}.

While bilinear interpolation can be used to resize the positional embeddings for a randomly selected patch size, resizing the patch embedding weights require more careful design choices. We mainly investigate two approaches in this work, namely bilinear interpolation (BL) and pseudo-inverse resize (PI-resize), which is inspired from FlexiViT. For a patch at different resolutions obtained through bilinear interpolation, the PI-Resize operator is designed to adjust the patch embedding weights to maximize the alignment of the extracted information. More formally, resizing a patch $x$ of size $p \times p$ into $\hat{x}$ of size $\hat{p} \times \hat{p}$ through bilinear interpolation could be represented as a linear transformation:
\begin{equation}
\hat{x} = B_{p}^{\hat{p}}vec(x)
\end{equation}
where $B_{p}^{\hat{p}} \in \mathbb{R}^{{\hat{p}}^2 \times p^2}$ is the bilinear interpolation matrix. Then, PI-Resize operator aims to obtain resized patch embedding weights $\hat{\omega}$ that satisfy the following optimization objective:
\begin{equation}
\hat{\omega} \in \arg \min _{\hat{\omega}} \mathbb{E}_{x \sim \mathcal{X}}\left[(\langle x, \omega\rangle-\langle \hat{x}, \hat{\omega}\rangle)^{2}\right]
\end{equation}
where $\langle a, b \rangle = vec(a)^{T}vec(b)$ and $\mathcal{X}$ is a distribution over the patches. For both of the upsampling ($\hat{p} \geq p$) and the downsampling ($\hat{p} < p$) cases, the $\hat{\omega}$ is recovered as:
\begin{equation}
\hat{\omega}=P_{p}^{\hat{p}}vec(\omega)
\end{equation}
where $P_{p}^{\hat{p}} \in \mathbb{R}^{\hat{p}^2 \times p^2}$ is the pseudoinverse of $({B_{p}^{\hat{p}}})^{T}$, thus the matrix representing the PI-Resize transformation. Although PI-Resize and resizing the spectrogram are related in function, they differ in that PI-Resize modifies the model parameters rather than the input spectrogram~\cite{beyer2022flexivit}. FlexiAST follows the same architecture and training process as standard ASTs after refining the weights of patch embeddings and positional embeddings for randomly selected patch sizes. In this work, supervised learning and PI-resize functions are employed as the default training and patch embedding weights resizing approaches respectively. These design choices are validated in Section~\ref{sec:ablation}.

\section{Experiments}

\subsection{Datasets and Evaluation Metrics}\label{sec:dataset}
\noindent\textbf{Datasets.} In our experiments, we train our method on (1) AudioSet Full, (2) VGGSound, (3) ESC-50, (4) Speech Commands, and (5) VoxCeleb datasets. 
 \textbf{AudioSet}~\cite{gemmeke2017audio} is a large-scale dataset that contains a diverse collection of audio samples, each labeled with a set of annotations. The dataset consists of over 2 million 10-second clips, covering a wide range of categories such as musical instruments, animal sounds, and human speech. There are a total of 527 labels. \textbf{VGGSound}~\cite{VGGSound} is a dataset that consists of nearly 200K 10-second videos. Each clip has been labeled with 309 sound classes, which include objects, human actions, and human-object interactions. \textbf{ESC-50}~\cite{piczak2015esc} dataset comprises 2,000 environmental audio recordings, each lasting for 5 seconds and classified into 50 categories. \textbf{Speech Commands-V2}~\cite{warden2018speech} contains \app 105K recordings, each lasting for 1 second, and featuring 35 commonly used speech commands. \textbf{VoxCeleb}~\cite{nagrani2020voxceleb} is an audio-visual human speech dataset with 1251 speakers and \app 145K utterances.

\noindent\textbf{Evaluation metrics.} Due to the presence of multiple labels in each sample in AudioSet, we evaluate it using the mean average precision (mAP) across all classes. In contrast to AudioSet, we report the Top-1 classification accuracy for the remaining datasets since only a single label is assigned for each sample.

\subsection{Implementation Details}
\noindent\textbf{Details of FlexiAST.} FlexiAST is architecturally identical to AST, with only differences in random patch size selection and patch resizing steps during training. Therefore, it has the same underlying patch size parameters as AST, which is set to 16x16 ($\omega$ for FlexiAST). The original AST provides additional techniques to further improve performance, such as patch overlapping, mixup augmentation~\cite{tokozumelearning}, padding, and model aggregation~\cite{izmailov2018averaging,breiman1996bagging}. For simplicity and to observe the direct impact of flexification, we do not use the above-mentioned settings in FlexiAST or standard ASTs presented in this paper. The rest of the default settings of AST are adapted identically. The original AST demonstrates that initialization enhances performance, and we also employ initialization in FlexiAST.  However, unlike standard AST, FlexiAST obtains initialization weights from standard ASTs such as AST-B/8, AST-B/16, or AST-B/32 instead of ViT~\cite{dosovitskiyimage} or DeiT~\cite{touvron2021training}. When initializing the weights of the FlexiAST, we adjust the patch embedding weights and positional embeddings of standard AST with PI-resize and bilinear interpolation respectively. In addition, we choose Supervised Training and PI-resize as the default options for the training paradigm and patch resizing operation, respectively, rather than Knowledge Distillation (KD) and bilinear interpolation (BL). The validation of these choices is discussed in Section~\ref{sec:ablation}.

\noindent\textbf{Details of Standard ASTs.} Unless specified otherwise, all standard ASTs (B/8, B/16, and B/32) within a particular dataset are trained for the same number of epochs, initialized with ViT, and learned using supervised learning with fixed patch sizes. However, for ESC-50 and Speech Commands ASTs, we initialize them from AudioSet pre-trained standard ASTs and then fine-tune them. After obtaining each standard AST in various datasets, AST-B/8 are used to initialize the FlexiAST. However, only for VoxCeleb dataset, AST-B/16 is employed.

\begin{table}
  \setlength{\tabcolsep}{0pt}
  \setlength{\extrarowheight}{5pt}
  \renewcommand{\arraystretch}{0.75}
  \newcolumntype{C}{>{\centering\arraybackslash}X}
  \newcolumntype{R}{>{\raggedleft\arraybackslash}X}
  \small
  \centering
  \scriptsize
  \begin{tabularx}{\linewidth}{p{1.8cm}p{0.01cm}Cp{0.01cm}Cp{0.01cm}Cp{0.01cm}Cp{0.01cm}Cp{0.01cm}Cp{0.01cm}Cp{0.01cm}Cp{0.01cm}Cp{0.01cm}C}
    \toprule[1pt]
\bf{Model} && \bf{48} && \bf{40} && \bf{32} && \bf{30} && \bf{24} && \bf{20} && \bf{16} && \bf{12} && \bf{10} && \bf{8} \\ 
    \midrule

    \multicolumn{21}{c}{\bf{VGGSound (Acc)}} \\
    \arrayrulecolor{lightgray}\midrule[0.25pt]\arrayrulecolor{black}
AST-B/8 &&0.003 &&0.004 &&0.005 &&0.006 &&0.012 &&0.033 &&0.080 &&0.221 &&0.363 &&0.500   \\ 
AST-B/16 &&0.009 &&0.023 &&0.073 &&0.105 &&0.234 &&0.328 &&0.502 &&0.300 &&0.132 &&0.013   \\ 
AST-B/32 &&0.121 &&0.284 &&0.455 &&0.366 &&0.237 &&0.119 &&0.030 &&0.008 &&0.005 &&0.003   \\ 
FlexiAST-KD &&0.362 &&0.415 &&0.448 &&0.453 &&0.464 &&0.476 &&0.487 &&0.487 &&0.493 &&0.493     \\
\rowcolor{lightgray!25}
FlexiAST-Sup. &&0.348 &&0.400 &&0.432 &&0.438 &&0.456 &&0.467 &&0.481 &&0.480 &&0.478 &&0.484   \\ 
    \arrayrulecolor{lightgray}\midrule[0.25pt]\arrayrulecolor{black}
        \multicolumn{21}{c}{\bf{AudioSet (mAP)}} \\
    \arrayrulecolor{lightgray}\midrule[0.25pt]\arrayrulecolor{black}
AST-B/8 &&0.006 &&0.006 &&0.006 &&0.006 &&0.007 &&0.009 &&0.025 &&0.151 &&0.276 &&0.397  \\ 
AST-B/16 &&0.008 &&0.011 &&0.028 &&0.044 &&0.172 &&0.297 &&0.396 &&0.248 &&0.100 &&0.024  \\ 
AST-B/32 &&0.074 &&0.211 &&0.371 &&0.348 &&0.181 &&0.069 &&0.021 &&0.010 &&0.007 &&0.006  \\ 
FlexiAST-KD &&0.306 &&0.337 &&0.353 &&0.362 &&0.378 &&0.387 &&0.394 &&0.399 &&0.397 &&0.397    \\
\rowcolor{lightgray!25}
FlexiAST-Sup. &&0.305 &&0.338 &&0.355 &&0.365 &&0.380 &&0.390 &&0.396 &&0.401 &&0.400 &&0.399   \\ 
    \arrayrulecolor{lightgray}\midrule[0.25pt]\arrayrulecolor{black}
    \multicolumn{21}{c}{\bf{ESC-50 (Acc)}} \\
    \arrayrulecolor{lightgray}\midrule[0.25pt]\arrayrulecolor{black}
AST-B/8  &&0.020 &&0.018 &&0.022 &&0.028 &&0.048 &&0.060 &&0.107 &&0.517 &&0.765 &&0.950  \\ 
AST-B/16 &&0.045 &&0.070 &&0.107 &&0.163 &&0.463 &&0.765 &&0.948 &&0.487 &&0.253 &&0.095  \\ 
AST-B/32 &&0.310 &&0.623 &&0.943 &&0.895 &&0.465 &&0.195 &&0.070 &&0.020 &&0.015 &&0.035  \\
FlexiAST-KD &&0.780 &&0.848 &&0.877 &&0.897 &&0.895 &&0.912 &&0.927 &&0.932 &&0.927 &&0.940    \\
\rowcolor{lightgray!25}
FlexiAST-Sup. &&0.752 &&0.870 &&0.882 &&0.892 &&0.910 &&0.920 &&0.930 &&0.943 &&0.930 &&0.940  \\  
    \arrayrulecolor{lightgray}\midrule[0.25pt]\arrayrulecolor{black}
    \multicolumn{21}{c}{\bf{Speech Commands (Acc)}} \\
    \arrayrulecolor{lightgray}\midrule[0.25pt]\arrayrulecolor{black}
AST-B/8 &&0.029 &&0.037 &&0.045 &&0.045 &&0.062 &&0.081 &&0.185 &&0.682 &&0.884 &&0.960  \\ 
AST-B/16 &&0.051 &&0.083 &&0.252 &&0.314 &&0.734 &&0.882 &&0.973 &&0.864 &&0.681 &&0.382  \\ 
AST-B/32 &&0.230 &&0.616 &&0.967 &&0.944 &&0.623 &&0.484 &&0.148 &&0.041 &&0.025 &&0.021  \\
FlexiAST-KD &&0.826 &&0.875 &&0.896 &&0.897 &&0.908 &&0.915 &&0.923 &&0.923 &&0.927 &&0.926  \\
\rowcolor{lightgray!25}
FlexiAST-Sup. &&0.823 &&0.893 &&0.921 &&0.929 &&0.944 &&0.951 &&0.960 &&0.962 &&0.964 &&0.964  \\  
    \bottomrule
  \end{tabularx}
    \caption{\textbf{Comparison of Standard ASTs and FlexiAST.}}\label{tab:exp_flexiall}
  \vspace{-0.9em}
\end{table}

\begin{figure}[t!]
\centering{
\resizebox{1.0\linewidth}{!}{%
\begin{tabular}{c}
\includegraphics[width =  1.0\linewidth]{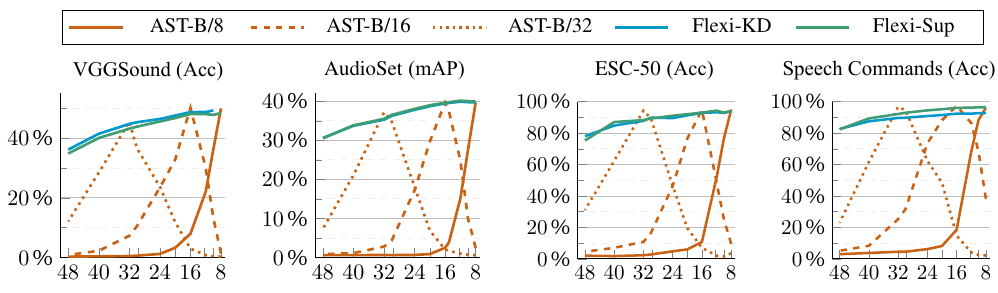} \\
\end{tabular}
}
}
\caption{\textbf{Visualization of Table~\ref{tab:exp_flexiall}.}} 
\label{fig:flexification_all}
\vspace{-2mm}
\end{figure}

\subsection{Results of Flexification}
\begin{figure}[t!]
\centering
{
\resizebox{\linewidth}{!}{%
\begin{tabular}{c}
\includegraphics[width = 1.0\linewidth]{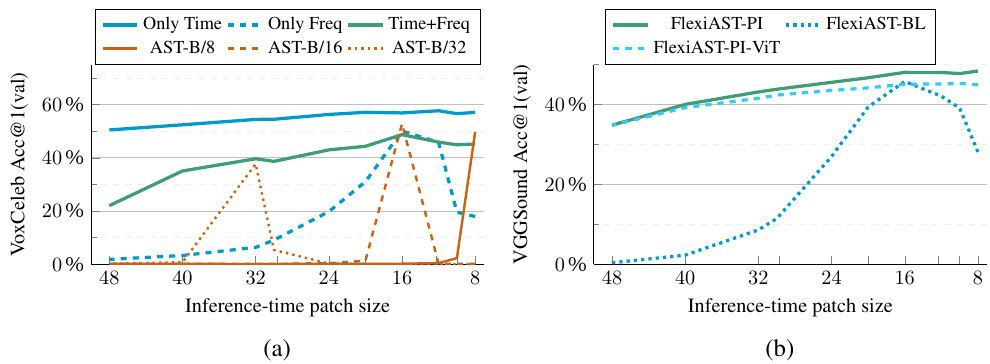} \\
\end{tabular}
}
}
\caption{\textbf{FlexiAST design choices on VoxCeleb and VGGSound.}}
\label{fig:voxceleb_vgg}
\vspace{-4mm}
\end{figure}

This section presents a comparison between our FlexiAST and the conventional ASTs trained using fixed patch sizes. We assess the performance of all models in inference stage across multiple patch sizes on five datasets, namely AudioSet, VGGSound, ESC-50, Speech Commands, and VoxCeleb. The evaluation results for first four dataset are presented in Table~\ref{tab:exp_flexiall} and Figure~\ref{fig:flexification_all}, and VoxCeleb results are displayed in Figure~\ref{fig:voxceleb_vgg}.

\noindent\textbf{Results for AudioSet, VGGSound, ESC-50 and Speech Commands.} Our results demonstrate that our FlexiAST displays a high degree of flexibility to handle a broad range of patch sizes during inference on all datasets, regardless of the training paradigm, whether it is \emph{Supervised} or \emph{Knowledge Distillation}. In contrast, the performance of standard ASTs deteriorates when evaluated on patch sizes different from those used during training. These results confirm that standard ASTs can be flexified with the proposed approach without any architectural changes. Notably, our FlexiAST model also performs comparably to the standard ASTs (or outperforms in some cases) when tested on the specific patch size they were trained on, such as 
FlexiAST-Sup. gives higher performance than AST-B/8 when evaluated at patch size 8 on AudioSet (Refer to Table~\ref{tab:exp_flexiall}). This concludes that our exploration provides a single model that can perform as good as a standard AST while retaining its ability to perform well when evaluated on various patch sizes.

\noindent\textbf{Results for VoxCeleb.} We find that when we apply our default flexification procedure (random patch size selection during training and resizing operation on patch embeddings), FlexiAST does not perform well on VoxCeleb dataset. The classification task in above mentioned datasets involves detecting the context of the samples, and so the resizing operation may not have a big impact since the context remains similar. However, the task of speaker identification in the VoxCeleb dataset requires speaker-specific information, in which the context of the sample is irrelevant, i.e., speaker embedding. Therefore, resizing patches in the frequency axis can directly affect the speaker-related information. To overcome this problem, we resize the patch size only in the time axis and apply resizing for the patch embedding weights accordingly during training. The result of this procedure is shown in Figure~\ref{fig:voxceleb_vgg} (a), demonstrating that this simple modification enables FlexiAST to work as expected and even perform beyond the standard ASTs. To further support this, we train FlexiAST by solely applying patch embedding resizing in the frequency axis. Results in Figure~\ref{fig:voxceleb_vgg} (a) demonstrate that resizing along the frequency axis has a significant harm on the flexibility of ASTs in speaker identification tasks.

\vspace{-2mm}
\subsection{Ablation Study}\label{sec:ablation}
We perform a series of ablative experiments to demonstrate the effectiveness of the techniques used in our proposed model and their alternatives. In order to reduce computational resources and time requirements, we mainly conduct the experiments on the VGGSound dataset. Key findings are as follows:

\noindent\textbf{1. Both supervised learning and knowledge distillation give good flexibility as a training paradigm.} We compare the performance of FlexiAST with different training paradigms. Specifically, we provide results in two settings: (1) Training with supervised learning, (2) Training with knowledge distillation (KD). For training with knowledge distillation, we use KL divergence as our loss and distill from the same standard AST model being used to initialize the FlexiAST model. The performance of these approaches is presented in Table~\ref{tab:exp_flexiall}. As depicted in the results, both training methodologies give similar performance generally and show marginal differences depending on the datasets, such as the Speech Commands dataset, where supervised learning shows a modest gap. However, due to its simplicity and shorter training time, we select the supervised learning as our default training method.

\noindent\textbf{2. PI-resize method provides flexibility while bilinear interpolation fails.} To demonstrate the impact of different resizing methods on FlexiAST, we utilize two distinct techniques, bilinear interpolation and PI-resize. In this experiment, we use FlexiASTs trained on VGGSound with supervised learning. The results of the experiment are presented in Figure~\ref{fig:voxceleb_vgg} (b), where we compare the effectiveness of the two methods. It is observed that while PI-resize makes successful flexification, bilinear interpolation (BL) fails to provide the desired flexibility.

\begin{figure}[t!]
\centering
{
\resizebox{\linewidth}{!}{%
\begin{tabular}{c}
\includegraphics[width = 0.4\linewidth]{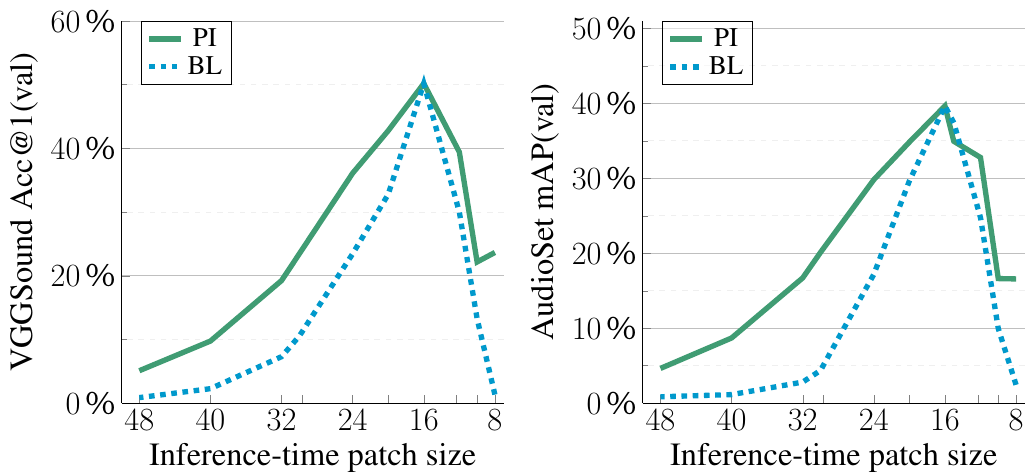} \\
\end{tabular}
}
}
\caption{\textbf{The effect of PI-resize in standard ASTs.} We evaluate the performance of Standard ASTs on various patch sizes with PI-resize and bilinear interpolation to see the impact of each resizing operation.}
\label{fig:BL_vs_PI}
\vspace{-2mm}
\end{figure}
\noindent\textbf{3. Utilizing PI-resize as a resizing technique to evaluate standard ASTs in various patch sizes helps for relatively better performance.}
The results of the previous experiment highlight that PI-resize is capable of providing flexibility to FlexiAST, whereas bilinear interpolation fails to do so. Additionally, standard ASTs are not flexible when they are evaluated on different patch sizes with bilinear interpolation (BL) as a conventional method. This leads us to question whether the use of PI-resize can improve the evaluation performance of standard ASTs on various patch sizes. Figure~\ref{fig:BL_vs_PI} presents comparison results obtained when PI-resize and BL are employed. Our findings indicate that standard ASTs do not perform well on patch sizes other than those they are trained on, regardless of the resizing method is used. However, PI-resize outperforms bilinear interpolation in terms of performance.

\noindent\textbf{4. Initialization from standard AST leads to a performance improvement.} We examine the impact of the source of initialization on FlexiAST in our experiment. We conduct two sets of experiments: (1) initializing FlexiAST from a standard AST trained on the same dataset using an 8x8 patch size, and (2) initializing FlexiAST from standard ViT. Figure~\ref{fig:voxceleb_vgg} (b) illustrates the experiment results that FlexiAST with standard AST initialization performs better than ViT initialization on the VGGSound dataset. However, It should be noted that even with the ViT initialization, our proposed method can still flexify.
\vspace{-2mm}\section{Conclusion}
In this paper, we focus on providing patch-size flexibility to ASTs. We identify that standard ASTs, which are trained with fixed patch sizes, lack flexibility, resulting in significant performance degradation when evaluated on various patch sizes that differ from the sizes they were trained on. To address this issue, we introduce a training approach that offers flexibility to existing ASTs without requiring any architectural changes. This is achieved by simply utilizing random patch size selection and resizing patch embedding weights accordingly. Additionally, we explore the necessity of task-specific resizing techniques, such as only resizing the time axis in speaker identification, to achieve optimal flexibility. As a result, our approach enables the creation of FlexiAST, a model that is versatile enough to handle all patch sizes with ease.
\newpage
\bibliographystyle{IEEEtran}
\bibliography{shortstrings,mybib}

\end{document}